\documentclass[conference]{IEEEtran}
\IEEEoverridecommandlockouts

    \usepackage[pdftex]{graphicx}

\usepackage{color,colortbl,multicol,amssymb,amsfonts,array}
\usepackage[cmex10]{amsmath}
\usepackage{gensymb,textcomp}
\usepackage[ansinew,utf8]{inputenc}
\usepackage[acronym,shortcuts,nomain,nopostdot]{glossaries}
\usepackage{blindtext}

\usepackage{booktabs}
\usepackage{dcolumn}

\newcolumntype{d}{D{.}{.}{-1}}
\newcolumntype{L}[1]{>{\raggedright\let\newline\\\arraybackslash\hspace{0pt}}m{#1}}
\newcolumntype{C}[1]{>{\centering\let\newline\\\arraybackslash\hspace{0pt}}m{#1}}
\newcolumntype{R}[1]{>{\raggedleft\let\newline\\\arraybackslash\hspace{0pt}}m{#1}}

\definecolor{to}{rgb}{0.8,0.8,0.8}
\definecolor{l1}{rgb}{0.95,0.95,0.95}
\definecolor{l2}{rgb}{0.92,0.92,0.92}

\newcommand{\lo}{\cellcolor{l1}}
\newcommand{\lt}{\cellcolor{l2}}
\newcommand{\tp}{\cellcolor{to}}
\newcommand{\tbt}[1]{\multicolumn{1}{c}{#1}}

\newcommand{\nal}{\multicolumn{1}{c|}{\textcolor[rgb]{0.75,0.75,0.75}{N/A}}}
\newcommand{\nag}{\multicolumn{1}{c}{\textcolor[rgb]{0.75,0.75,0.75}{\cellcolor{l2}N/A}}}
\newcommand{\nagl}{\multicolumn{1}{c|}{\textcolor[rgb]{0.75,0.75,0.75}{\cellcolor{l2}N/A}}}
\newcommand{\lona}{\multicolumn{1}{c}{\textcolor[rgb]{0.75,0.75,0.75}{\cellcolor{l1}N/A}}}
\newcommand{\ze}{\multicolumn{1}{d}{\color[rgb]{0.75,0.75,0.75}0}}
\newcommand{\zeg}{\multicolumn{1}{d}{\lt\color[rgb]{0.75,0.75,0.75}0}}
\newcommand{\zel}{\multicolumn{1}{d|}{\color[rgb]{0.75,0.75,0.75}0}}
\newcommand{\zegl}{\multicolumn{1}{d|}{\lt\color[rgb]{0.75,0.75,0.75}0}}

\newcommand{\lu}[1]{log$_{\text{10}}$#1} 

\definecolor{hyp}{rgb}{0,0,0.3} 
\usepackage[colorlinks=true,linkcolor=hyp,filecolor=hyp,citecolor=hyp,urlcolor=hyp]{hyperref}

\newglossarystyle{acro}{
    \glossarystyle{super}

}
\newcommand{\acro}[2]{\newacronym{#1}{#1}{#2}}      
\newcommand{\acrol}[3]{\newacronym{#1}{#2}{#3}}     

\newcommand{\acrop}[4]{\newacronym[plural=#3,firstplural=#4 (#3)]{#1}{#1}{#2}}

\newcommand{\acF}[1]{\acf{#1}\glsunset{#1}}

\newcommand{\acFp}[1]{\glsreset{#1}\acp{#1}}

\newcommand{\acX}[1]{\acrlong{#1}}

\acro{3GPP}{3rd generation partnership project}
\acrol{1-D}{\mbox{1-D}}{one-dimensional}
\acrol{2-D}{\mbox{2-D}}{two-dimensional}
\acrol{3-D}{\mbox{3-D}}{three-dimensional}
\acrol{4-D}{\mbox{4-D}}{four-dimensional}
\acrol{6-D}{\mbox{6-D}}{six-dimensional}
\acro{3G}{third generation}
\acro{4G}{fourth generation}
\acro{5G}{fifth generation}

\acrop{AoA}{azimuth angle of arrival}{AoAs}{azimuth angles of arrival}
\acrop{AoD}{azimuth angle of departure}{AoDs}{azimuth angles of departure}
\acro{ASA}{azimuth spread of arrival}
\acro{AS}{angular spread}
\acro{ASD}{azimuth spread of departure}
\acro{AWGN}{additive white Gaussian noise}
\acro{AGC}{adaptive gain control}
\acro{ACF}{autocorrelation function}
\acro{ASE}{average squared error}

\acro{BD}{block diagonalization}
\acro{BS}{base station}
\acro{BC}{broadcast channel}
\acro{BPSK}{binary phase-shift keying}
\acro{BP}{break point}

\acro{CDF}{cumulative distribution function}
\acro{CIR}{channel impulse response}
\acro{CoMP}{coordinated multi-point}
\acro{CoF}{cost of feedback}
\acro{CRONOS}{cellular radio network simulator}
\acro{CSI}{channel state information}
\acro{COST}{European Cooperation in Science and Technology}
\acro{CPS}{cyber physical system}
\acro{CL}{clutter loss}

\acro{DS}{delay spread}
\acro{DD}{Doppler-delay}
\acro{DFT}{discrete Fourier transform}
\acro{DPC}{dirty-paper coding}
\acro{D2D}{device-to-device}

\acro{ESA}{elevation spread of arrival}
\acro{ESD}{elevation spread of departure}
\acrop{EoA}{elevation angle of arrival}{EoAs}{elevation angles of arrival}
\acrop{EoD}{elevation angle of departure}{EoDs}{elevation angles of departure}

\acro{FBR}{feedback rate}
\acro{FBS}{first-bounce scatterer}
\acro{FI}{feedback interval}
\acro{FR}{frequency response}
\acro{FR1}{frequency reuse 1}
\acro{FR4}{frequency reuse 4}
\acro{FIR}{finite impulse response}
\acro{FFT}{fast Fourier transform}
\acro{FWHM}{full width at half maximum}
\acro{FDD}{frequency-division duplexing}
\acro{FSPL}{free-space path loss}

\acro{GCS}{global coordinate system}
\acro{GF}{geometry factor}
\acro{GR}{ground reflection}
\acro{GoC}{gain of CoMP}
\acro{GoP}{gain of prediction}
\acro{GPS}{global positioning system}
\acro{GSCM}{geometry-based stochastic channel model}

\acro{HHI}{Heinrich Hertz Institute}
\acro{HPBW}{half-power beamwidth}
\acro{HAP}{high-altitude platform}

\acro{ISD}{inter site distance}
\acrol{iid}{i.i.d.}{independent and identically distributed}
\acro{IDFT}{inverse discrete Fourier transform}
\acro{IFFT}{inverse fast Fourier transform}
\acro{IR}{impulse response}
\acro{ITU}{International Telecommunication Union}
\acro{IF}{intermediate frequency}

\acro{JT}{joint transmission}
\acro{JCF}{joint correlation function}

\acro{KF}{Ricean K-factor}
\acro{KPI}{key performance indicator}

\acro{LBS}{last-bounce scatterer}
\acro{LHCP}{left hand circular polarized}
\acro{LOS}{line of sight}
\acro{LSP}{large-scale parameter}
\acro{LTE}{long term evolution}
\acro{LSAS}{large-scale antenna systems}
\acro{LSF}{large-scale fading}
\acro{LEO}{low Earth orbit}

\acro{MET}{multi-user eigenmode transmission}
\acro{MIMO}{multiple-input multiple-output}
\acro{MISO}{multiple-input single-output}
\acro{MMSE}{minimum mean square error}
\acro{MPC}{multipath component}
\acro{MS}{mobile station}
\acro{MSE}{mean square error}
\acro{MT}{mobile terminal}
\acro{MU-MIMO}{multi-user MIMO}
\acro{MIMOSA}{MIMO over satellite}
\acro{MAC}{multiple-access channel}
\acro{MRT}{maximum ratio transmission}

\acro{NLOS}{non-line of sight}
\acro{NR}{new radio}
\acro{NTN}{non-terrestrial network}

\acro{OFDM}{orthogonal frequency division multiplexing}
\acro{O2I}{outdoor-to-indoor}
\acro{ODA}{omni directional array}

\acro{PDP}{power delay profile}
\acro{PDF}{probability density function}
\acro{PG}{path gain}
\acro{PL}{path loss}
\acro{PUCA}{polarized uniform circular array}
\acro{PAS}{power-angular spectrum}
\acro{P2P}{peer-to-peer}
\acro{PSCS}{Propsound channel sounder}

\acro{QAM}{quadrature amplitude modulation}
\acro{QuaDRiGa}{quasi deterministic radio channel generator}
\acro{QRDRLS}{QR-decomposition recursive least squares}
\acro{QoS}{quality of service}

\acro{RB}{resource block}
\acro{RHCP}{right hand circular polarized}
\acro{RLS}{recursive least-squares}
\acro{RLE}{run-length encoding}
\acro{RX}{receiver}
\acro{RMS}{root mean square}
\acro{RAN}{radio access network}
\acro{RSRP}{reference signal received power}
\acro{RF}{radio frequency}

\acro{SAGE}{space-alternating generalized expectation-maximization}
\acro{SCM}{spatial channel model}
\acro{SF}{shadow fading}
\acro{SV}{singular value}
\acro{SVD}{singular value decomposition}
\acro{SNR}{signal to noise ratio}
\acro{SIR}{signal to interference ratio}
\acro{SINR}{signal to interference and noise ratio}
\acro{SISO}{single input single output}
\acro{SIMO}{single input multiple output}
\acro{STD}{standard deviation}
\acro{SSG}{state sequence generator}
\acro{SSF}{small-scale-fading}
\acro{SOS}{sum-of-sinusoids}
\acro{SC}{spatial consistency}

\acro{TP}{throughput}
\acro{TUB}{Technische Universit\"{a}t Berlin}
\acro{TX}{transmitter}

\acro{ULA}{uniform linear array}
\acro{UML}{unified modeling language}
\acro{UMi}{urban-microcell}
\acro{UMa}{urban-macrocell}
\acro{UAV}{unmanned aerial vehicle}
\acro{UE}{user equipment}

\acro{VR}{visibility region}

\acro{WINNER}{Wireless World Initiative for New Radio}
\acro{WSSUS}{wide sense stationary uncorrelated scattering}
\acro{WSS}{wide-sense stationary}
\acro{WGS}{world geodetic system}

\acro{XPD}{cross-polarization discrimination}
\acro{XPR}{cross polarization ratio}

\acro{ZF}{zero-forcing}
\acro{ZoA}{zenith angle of arrival}
\acro{ZoD}{zenith angle of departure}
\acro{ZSA}{zenith angle spread of arrival}
\acro{ZSD}{zenith angle spread of departure}
\acro{ZCR}{zero-crossing rate}

\newcolumntype{L}[1]{>{\raggedright\let\newline\\\arraybackslash\hspace{0pt}}m{#1}}
\newcolumntype{C}[1]{>{\centering\let\newline\\\arraybackslash\hspace{0pt}}m{#1}}
\newcolumntype{R}[1]{>{\raggedleft\let\newline\\\arraybackslash\hspace{0pt}}m{#1}}

\begin{document}

\title{A 5G-NR Satellite Extension for the QuaDRiGa Channel Model\\
\thanks{\textbf{\textsc{Acknowledgement:}} The research leading to these results has received funding from the European Union H2020 5GPPP under grant n. 815323 and supported by the Institute for Information \& communications Technology Promotion (IITP) grant funded by the Korea government (MSIT) (No.2018-0-00175, 5G AgiLe and fLexible integration of SaTellite And cellulaR).}
}

\author{\IEEEauthorblockN{ Stephan Jaeckel\IEEEauthorrefmark{1}, Leszek Raschkowski\IEEEauthorrefmark{2}, and Lars Thiele\IEEEauthorrefmark{2}}
\IEEEauthorblockA{\IEEEauthorrefmark{1} SJC Wireless, Berlin, Germany, jaeckel@sjc-wireless.com}\IEEEauthorblockA{\IEEEauthorrefmark{2}Fraunhofer Heinrich Hertz Institute, Berlin, Germany}}

\maketitle

\begin{abstract}
\Acf{LEO} satellite networks will become an integral part of the global telecommunication infrastructure. Modeling the radio-links of these networks and their interaction with existing terrestrial systems is crucial for the design, planning and scaling of these networks. The \ac{3GPP} addressed this by providing guideline for such a radio-channel model. However, the proposed model lacks a satellite orbit model and has some inconsistencies in the provided parameters. This is addressed in this paper. We provide a non-geostationary-satellite model that can be integrated into \acp{GSCM} such as QuaDRiGa. We then use this model to obtain the \ac{GSCM} parameters from a simplified environment model and compare the results to the 3GPP parameter-set. This solves the inconsistencies, but our simplified approach does not consider many propagation effects. Future work must therefore rely on measurements or accurate Ray-tracing models to obtain the parameters.
\end{abstract}

\section{Introduction}

3GPP released a comprehensive study on \acFp{NTN} \cite{3gpp_tr_38811_v1520} in order include space or airborne vehicles into the 5G infrastructure. These offer wide service coverage capabilities and reduced vulnerability to physical attacks or natural disasters. The idea is to foster the roll-out of 5G services in unserved areas, reinforce the 5G service reliability, and enable improved network scalability. To enable simulation studies, such as link-budget analysis, link and system-level performance studies or coexistence analysis with terrestrial cellular networks, channel model guidelines have been provided. A model calibration was done in \cite{3gpp_tr_38821_v1600} for several aspects of the model. To support these activities, \acp{NTN} have been added to the QuaDRiGa channel model \cite{quadriga_www}. However, since QuaDRiGa has been developed primarily for terrestrial applications, some modifications are necessary to incorporate \acp{NTN}. On the other hand, many additional modeling components are already available that go beyond the 3GPP guidelines. This allows more complex simulations to be conducted, but requires modifications to the model:

\paragraph{Coordinate system} The 3GPP \ac{NTN} model \cite{3gpp_tr_38811_v1520} uses a simplified ''Earth centered Earth fixed'' coordinate system, whereas the terrestrial 3GPP model \cite{3gpp_tr_38901_v1610} uses metric local Cartesian coordinates. In order to combine the two models, we provide a coordinate transformation that maps orbital positions and trajectories into the local Cartesian coordinates.

\paragraph{Frequency range} \cite{3gpp_tr_38901_v1610} includes an optional model for multiple frequencies which has been implemented in QuaDRiGa \cite{Jaeckel2019}. However, this option was not considered by \cite{3gpp_tr_38811_v1520}. Hence, \cite{3gpp_tr_38811_v1520} only provides model parameters for the S-band (2-4 GHz) and for the KA-band (26.5-40 GHz). We provide parameters that support a continuous frequency range from 2-40 GHz. Thus, we can use the multi-frequency option and also perform simulations at the commonly used KU-band (10.7-17.5 GHz).

\paragraph{Spatial Consistency} Another optional feature provided by \cite{3gpp_tr_38901_v1610} is \ac{SC} which solves the problem of achieving realistic correlations in multi-user wireless channels. This becomes important for multi-satellite simulations where, for example, the \ac{LOS} to multiple satellites might be blocked by the same building. \Ac{SC} is available in QuaDRiGa \cite{Jaeckel2018} and can be used for \ac{NTN} channels.

\paragraph{Mobility} Satellites in \acX{LEO} are highly mobile, causing large differential delays and Doppler shifts. However, none of the 3GPP models supports mobility at both ends of the link. The assumption in \cite{3gpp_tr_38811_v1520} is that all satellite positions are fixed. A dual-mobility model is available in QuaDRiGa \cite{Jaeckel2019,Jaeckel2018}. It can be used to simulate entire satellite constellations and track orbital movements for a longer time-period. This requires that the model parameters are given as a function of the satellite elevation angle. 

\Acp{GSCM} have of two main components: a stochastic part that generates a random propagation environment around the \ac{MT} location on Earth, and a deterministic part that lets transmitters (e.g., the satellites) and receivers (the \acp{MT}) interact with this environment. The stochastic part requires model parameters to be extracted from radio channel measurements. However, in order to capture all model parameters, such measurements require specific channel sounding hardware. Channel measurements using satellites (e.g.\ \cite{Eberlein2011}) are then either limited in bandwidth, elevation angle range, and spatial resolution, or they are done with terrestrial transceivers (e.g.\ \cite{Burkhardt2014}). Another approach is to use deterministic Raytracing simulations to obtain the \ac{GSCM} parameters. This has been favored by the 3GPP community. However, the simulations for \cite{3gpp_tr_38811_v1520} were done using \acp{HAP} and there are inconsistencies in the environments. This leads to questionable results when comparing the parameters of different environments, such as the Ricean K-Factor for the Suburban and Rural scenarios. For this reason, we used a different approach where we generate random satellite constellations, \ac{MT} positions and propagation environments. Based on these inputs, radio channel coefficients are generated in a purely deterministic way using QuaDRiGa. The data is then analyzed in the same way as measurement or Raytracing data would be. In this way, we get a complete set of consistent parameters for the stochastic model. 

The paper is structured as follows: Section~\ref{sec:non-gso-model} introduces the satellite orbit model and the coordinate transformation procedure. Section~\ref{sec:lsf-model} then combines this model with the existing radio channel model to obtain the \ac{LSF} parameters. Results are discussed in Section~\ref{sec:lsf-model-results}. An open-source implementation is available as part of the \acf{QuaDRiGa} \cite{quadriga_www}.

\section{Non-GSO Satellite Orbit Model}\label{sec:non-gso-model}

\begin{figure}[hbt]
    \centering
    \includegraphics[width=86mm]{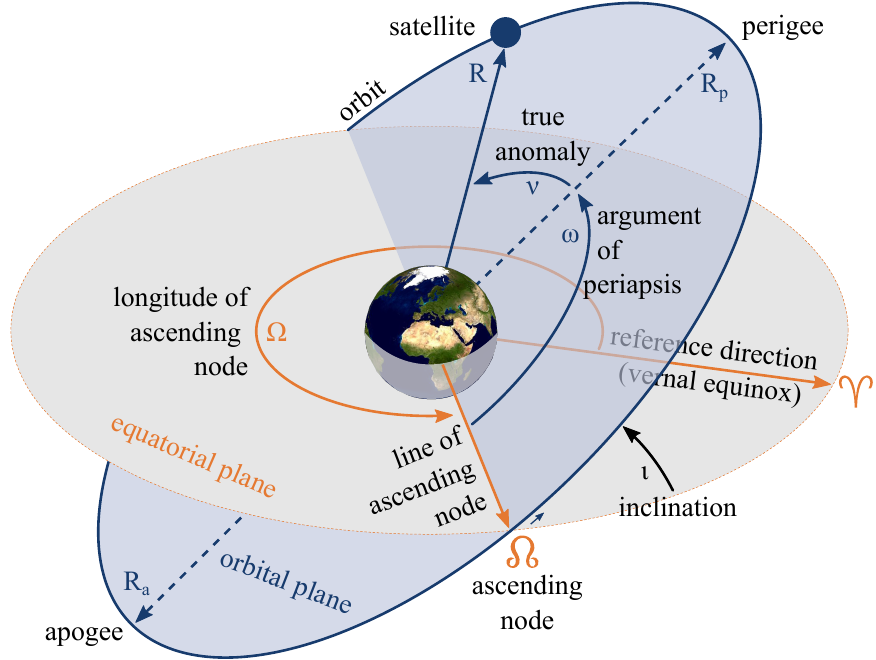}
    \vspace{-0.5\baselineskip}
    \caption{Diagram illustrating various terms in relation to satellite orbits}
    \label{fig:orbit_terms}
\end{figure}

\paragraph{Orbit model}
The satellite orbit model \cite{itu-r_s1503-3_2018} uses Earth's attraction as the main factor for orbital motion. Six parameters define the satellite position (see Fig.~\ref{fig:orbit_terms}): 1) the length of the semi-major axis $a=\frac{R_a+R_p}{2}$; 2) the eccentricity $e=\frac{R_a-R_p}{R_a+R_p}$ determines the amount by which an orbit deviates from a circle (0 yields a circular orbit); 3) the inclination angle $\iota$ measures the tilt of the orbit; 4) the longitude of the ascending node $\Omega$ orients the point where the orbit passes upward through the equatorial plane; 5) the argument of periapsis $\omega$ defines the orientation of the ellipse in the orbital plane; and, 6) the true anomaly $\nu$ defines the position of the satellite along that ellipse. Given the values $\Omega_0$, $\omega_0$ and $\nu_0$ at a reference time, orbit mechanics predict the position of the satellite in the future. 

\begin{table}[h]
    \vspace{-\baselineskip}
    \setlength{\tabcolsep}{3pt}
    \scriptsize
    \centering
    \caption{Constants required for orbit prediction}\label{tab:Constants}
    \vspace{-2mm}
\begin{tabular}{ l | c | c  | c  }
Parameter & Notation & Value & Unit \\
\hline\hline
Radius of the Earth               & $R_e$      & $6378.137$                         & km\\
Mass of the Earth                 & $M_e$      & $5.9722 \cdot 10^{24}$             & kg\\
Earth' rotation period            & $T_e$      & $86164.09054$                      & s \\
Earth's angular rotation rate     & $\omega_e$ & $7.29211585453 \cdot 10^{-5} $     & rad/s \\
Gravitational constant            & $G$        & $6.67408 \cdot 10^{-20}$           & $\text{km}^3$/$\text{s}^2$/$\text{kg}$ \\
Earth's non-sphericity factor     & $J_2$      & $0.001082636$                      & - \\
\end{tabular}

\end{table}

Orbit perturbations are mainly due to Earth's oblateness. This is modeled by changing the ascending node longitude and perigee argument. For a given time point $t$ relative to the reference time, the values $\Omega(t)$ and $\omega(t)$ are updated to
\begin{eqnarray}
    \Omega(t) &=&  \Omega_0 - t \cdot \bar{n} \cdot \bar{p} \cdot \cos \iota,  \\
    \omega(t) &=&  \omega_0 + t \cdot \bar{n} \cdot \bar{p} \cdot \left( 2 - 2.5 \cdot \sin^2 \iota \right),
\end{eqnarray}
where the parameters $\bar{n}$ and $\bar{p}$ are given by
\begin{equation}
 \bar{n} = \sqrt{\frac{G\cdot M_e}{a^3}} \cdot \left( 1+\bar{p} \cdot \left( 1-1.5 \cdot \sin^2 \iota \right) \cdot \sqrt{1-e^2} \right), 
\end{equation}
\begin{equation}
 \bar{p} = \frac{3 \cdot J_2 \cdot R_e^2}{2a^2\cdot (1-e^2)^2}.
\end{equation}
The constants $M_e$, $R_e$, $G$ and $J_2$ can be found in Table~\ref{tab:Constants}. An update of $\nu(t)$ is calculated using the eccentric anomaly $\left\{E_0, E(t)\right\}$ instead of the true anomaly $\left\{\nu_0, \nu(t)\right\}$ by solving
\begin{equation}
     E(t) - e \cdot \sin E(t) = E_0 - e \cdot \sin E_0 + \bar{n} \cdot t,
\end{equation}
where the transformation between $E$ and $\nu$ follows from 
\begin{equation}
     \tan \left( \frac{E}{2} \right) = \sqrt{ \frac{1+e}{1-e} } \cdot \tan \left( \frac{\nu}{2} \right).
\end{equation}
With the updated parameters $\Omega(t)$, $\omega(t)$ and $\nu(t)$ it is possible to calculate the satellite position in Cartesian coordinates by
\begin{equation}
      x_i = R \cdot \left\{ \cos(\omega+\nu) \cdot \cos\Omega - \sin(\omega+\nu) \cdot \sin\Omega \cdot \cos\iota \right\}
\end{equation}
\begin{equation}
      y_i = R \cdot \left\{ \cos(\omega+\nu) \cdot \sin\Omega - \sin(\omega+\nu) \cdot \cos\Omega \cdot \cos\iota \right\}
\end{equation}
\begin{equation}
      z_i = R \cdot \sin(\omega+\nu) \cdot \sin\iota,
\end{equation}
where $R$ is the distance between Earth's center and the satellite
\begin{equation}
     R(t) = \frac{a\cdot(1-e)^2}{1+e\cdot \cos \nu(t) }.
\end{equation}
To calculate the satellite coordinates as seen by an observer on Earth, Earth's rotation needs to be taken into account. This is done by translating the satellite positions into a geographic coordinate system and adding the Earth's angular rotation.
\begin{eqnarray}
      \theta_r(t) &=& \arctan_2\left\{ z_i(t), \sqrt{x_i^2(t)+y_i^2(t)} \right\}\\
      \phi_r(t) &=&  \arctan_2\left\{ y_i(t), x_i(t) \right\} - \omega_e \cdot t
\end{eqnarray}
$\arctan_2(y,x)$ is the four quadrant inverse tangent of the elements $y$ and $x$ having values between $-\pi$ and $\pi$. At the reference time $t=0$, Earth's prime meridian is aligned with the vernal equinox. The satellite coordinates in rotating Cartesian coordinates $(x_r,y_r,z_r)$ follow from the transformation
\begin{eqnarray}
  \label{eq:sph2cartx}    x &=& R \cdot \cos\phi \cdot \cos\theta, \\
  \label{eq:sph2carty}    y &=& R \cdot \sin\phi \cdot \cos\theta, \\
  \label{eq:sph2cartz}    z &=& R \cdot \sin\theta.
\end{eqnarray}

\paragraph{Coordinate transformation}
The \ac{MT}-centric coordinate system is defined by a tangential plane having its origin at a reference position on Earth given by its longitude $\phi_u$, latitude $\theta_u$ and radius $R_e$. The transformation \eqref{eq:sph2cartx}-\eqref{eq:sph2cartz} converts these to Cartesian coordinates $(x_u,y_u,z_u)$. The transformation of the satellite coordinates $(x_r,y_r,z_r)$ into \ac{MT}-centric coordinates is done by 
\begin{equation}
   \left[ \begin{array}{c}
   x_q(t) \\ 
   y_q(t) \\ 
   z_q(t)
   \end{array} \right]
   = \mathbf{R}_q \cdot 
   \left[ \begin{array}{c}
   x_r(t) - x_u \\ 
   y_r(t) - y_u\\ 
   z_r(t) - z_u
   \end{array} \right],
\end{equation}
where the rotation matrix $\mathbf{R}_q$ aligns the geographic Cartesian coordinate system with the \ac{MT}-centric coordinate system whose $x$-axis points eastwards and $y$-axis points northwards.
\begin{equation}
\mathbf{R}_q = 
   \left[ \begin{array}{ccc}
   -\sin\phi_u               & \cos\phi_u                & 0 \\ 
   -\sin\theta_u \cos\phi_u  & -\sin\theta_u \sin\phi_u  & \cos\theta_u \\ 
   \cos\theta_u \cos\phi_u   & \cos\theta_u \sin\phi_u   & \sin\theta_u
   \end{array} \right]
\end{equation}
The satellite is visible above the horizon when $z_q(t)>0$ and its elevation angle is
\begin{equation}
  \alpha(t) = \arctan_2\left\{ z_q(t), \sqrt{x_q^2(t)+y_q^2(t)} \right\}.
\end{equation}
Satellites use directional antennas. Hence, the satellite's orientation towards the observer on Earth is important. The following steps calculate this  orientation, assuming that the satellite is spinning at one revolution per orbit so that the same side always faces the Earth. First, three vectors are calculated 
\begin{eqnarray}
      \mathbf{U} &=& \left[ \begin{array}{cccc}
           x_u & y_u & z_u
           \end{array} \right]^T,\\
      \mathbf{R}(t) &=& \left[ \begin{array}{ccc}
           x_r(t) & y_r(t) &  z_r(t)
           \end{array} \right]^T, \\
      \mathbf{D}(t) &=& \mathbf{R}(t+\Delta t) - \mathbf{R}(t).
\end{eqnarray}
They are normalized to unit-length vectors $\bar{\mathbf{U}}$, $\bar{\mathbf{R}}(t)$ and $\bar{\mathbf{D}}(t)$. The vector $\bar{\mathbf{D}}(t)$ is the direction of travel calculated from two orbital positions at time points $t$ and $t+\Delta t$. The \textit{bank angle} $\beta$ is the orientation around the axis drawn through the body of the satellite from tail to nose, relative to the tangential plane.
\begin{equation}
    \beta_q(t) = \arcsin\left\{ \bar{\mathbf{U}}^T \cdot \left( \bar{\mathbf{R}}(t) \times \bar{\mathbf{D}}(t) \right)  \right\}.
\end{equation}
The \textit{heading angle} $\gamma$ is the pointing direction of the satellite.
\begin{eqnarray}
        \bar{\mathbf{D}}_q(t) &=& \mathbf{R}_q  \cdot \bar{\mathbf{D}}(t)\\
        \gamma_q(t) &=& \arctan_2\left\{ y_{\bar{\mathbf{D}}_q}(t), x_{\bar{\mathbf{D}}_q}(t) \right\}
\end{eqnarray}
The \textit{tilt angle} $\delta$ is the vertical orientation of the satellite.
\begin{equation}
    \delta_q(t) = \arctan_2\left\{ z_{\bar{\mathbf{D}}_q}(t), \sqrt{x_{\bar{\mathbf{D}}_q}^2(t)+y_{\bar{\mathbf{D}}_q}^2(t)} \right\}
\end{equation}
The six parameters $x_q,y_q,z_q,\beta_q,\gamma_q,\delta_q$ define the satellite's position and orientation as seen by an observer on Earth. Hence, satellites can be used as transmitters in \acp{GSCM} such as QuaDRiGa. Their orbital motion can be tracked over time and so can be their communication links. This enables the realistic simulation of the propagation channels of entire satellite networks.

\section{Obtaining GSCM Model Parameters}\label{sec:lsf-model}

\begin{figure}[b]
    \centering
    \vspace{-\baselineskip}
    \includegraphics[width=82mm]{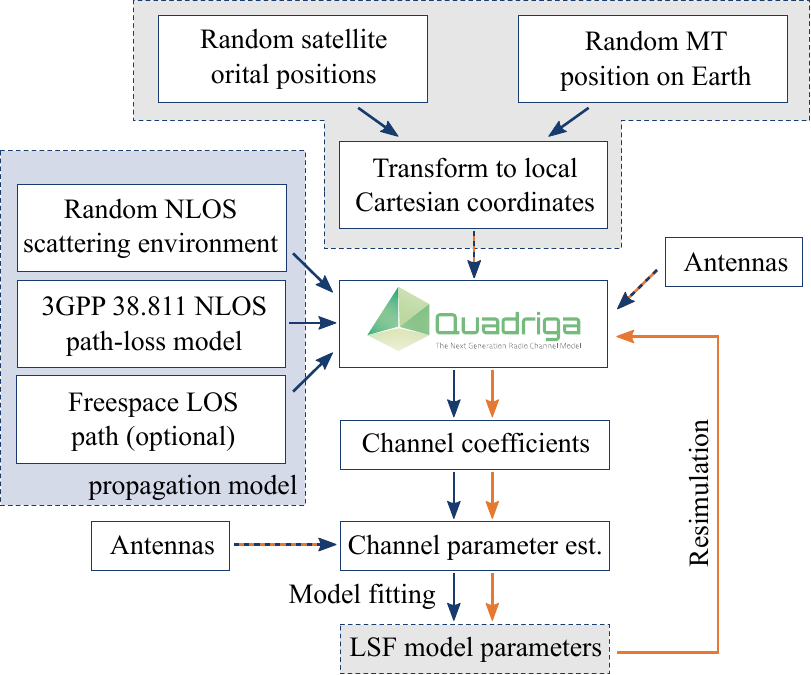}
    \vspace{-0.5\baselineskip}
    \caption{Satellite LSF model parameter estimation workflow}
    \label{fig:model_parameter_workflow}
\end{figure}

\Acp{GSCM} are parametric models, i.e.\ the properties of the communication links depend on the model parameters. In this section, we describe a procedure to obtain these parameters. An overview is given in Fig.~\ref{fig:model_parameter_workflow}. We created random satellite constellations using Walker-Delta patterns at three different orbit heights: 550~km, 2,000~km, and 20,200~km. The inclination angles were 53\degree , 61\degree, and 63\degree , respectively. Random \ac{MT} positions were chosen in between $\pm$53\degree\ latitude on Earth, assuming that the \ac{MT} is outdoors at 1.5~m height above ground. The positions were imported into the QuaDRiGa channel model using the coordinate transformation from Sec.~\ref{sec:non-gso-model}. Then, we created a simplified random propagation environment around the MT position. The assumption is that in a satellite channel, \ac{NLOS} paths must come from objects close to the \ac{MT} (e.g., buildings, trees, cars, etc.). The distribution of these objects depends on the scenario. Seen from the \ac{MT}, we created between 6 and 10 random arrival directions drawn from a Uniform distribution in the range $[-\pi,\pi[$. The distances to the scatterers are modeled by a truncated Gaussian distribution having a mean and \ac{STD} according to Table~\ref{tab:propagation_environment}. Truncation was done at the minimum and maximum values in Table~\ref{tab:propagation_environment}. Those distances reflect the building density in the environment. 

A second parameter describes the height of the scatterers above the ground. This parameter reflects the typical building heights in the environment. Most scattered paths come from objects at the same height as the \ac{MT}. However, high buildings in urban and dense urban settings also cause paths arriving from higher elevation angles. For the sake of simplicity, only single-bounce scattering is assumed. Multiple reflections of the signal would not change the arrival \ac{AS} and the effect on the departure \ac{AS} is negligible due to the large distance to the satellite in orbit. We further assume that each \ac{NLOS} path carries on average the same amount of power. Hence, for each satellite-\ac{MT} link, we can obtain a value for the \ac{NLOS} \ac{PL} and \ac{SF} according to \cite{3gpp_tr_38811_v1520} and divide this value by the number of \ac{NLOS} paths in the scenario. For the \ac{LOS} model, we simply add a free space path to the \ac{NLOS} model. Departure and arrival direction are defined by the satellite and \ac{MT} positions and the power is given by the \ac{FSPL} model.

\begin{table}[t]
    \setlength{\tabcolsep}{3pt}
    \scriptsize
    \centering
    \caption{Propagation environment parameters}\label{tab:propagation_environment}
    \vspace{-2mm}
\begin{tabular}{ l | c | cccc | cccc   }
Scenario & no. & \multicolumn{4}{c|}{Distance to scatterer (m)} & \multicolumn{4}{c}{Scatterer height (m)} \\
& paths & min. & max. & mean & STD & min. & max. & mean & STD \\
\hline\hline
Dense Urban & 10 & 0.1 &  100 &  40 &  30 & 0 & 60 &   2 &  18 \\
Urban       & 10 & 0.1 &  200 &  50 &  35 & 0 & 30 &   2 &   9 \\
Suburban    &  8 & 0.1 &  500 &  65 &  50 & 0 &  8 & 1.5 & 1.5 \\
Rural       &  6 & 0.1 & 3500 & 300 & 200 & 0 &  8 & 1.5 & 1.5 \\
\end{tabular}
\vspace{-\baselineskip}
\end{table}

To obtain channel coefficients from QuaDRiGa, we also need antenna models for the satellite and the \ac{MT}. For the satellite, we used a simple omnidirectional \ac{LHCP} antenna. However, at the \ac{MT}, we want to estimate the arrival angles from the channel coefficients. Hence, we need an array antenna with sufficient spatial resolution. This is achieved by placing 28 vertically polarized patch elements around a sphere. The distance of each element to the center of the sphere is 1.6~$\lambda$ and each pair of neighboring elements is placed 1~$\lambda$ apart. Side lobes are suppressed by reducing the \ac{FWHM} of the single elements to 20\degree\ . This ideal spherical array antenna has a gain of 14 dBi and a \ac{FWHM} of 16\degree\ for all arrival directions. Using the algorithms from \cite{Jaeckel2017_diss}, it is possible to estimate the arrival direction of a path in azimuth and elevations with less than 1\degree\ error. Additional circularly polarized elements are used to calculate the \ac{XPR}.

The remaining steps of the procedure are the generation of the channel coefficients, the extraction of the channel parameters (delays, angles, \ac{DS}, \acp{AS}, K-Factor, etc.) and fitting a multilinear regression model to these parameters. The complete list of parameters consists of
\begin{itemize}
    \item the \acF{PL} and \acF{SF},
    \item the \acF{KF},
    \item the \acF{RMS} \acF{DS},
    \item the \ac{RMS} \acF{ASA},
    \item the \ac{RMS} \acF{ASD},
    \item the \ac{RMS} \acF{ESA},
    \item the \ac{RMS} \acF{ESD}, and
    \item the \acF{XPR}.
\end{itemize}
The methods for extracting the channel parameters are described in \cite{Jaeckel2017_diss}. The linear model for the \acp{LSP} is given by
\begin{multline}\label{eq:lsf_linear_model}
    V = V_\mu + V_\epsilon \cdot  \log_{10} d+ V_\gamma \cdot \log_{10} f_\text{GHz}  + 
        V_\alpha \cdot  \log_{10}\alpha_\text{rad} + \\
        X \left( {V}_\sigma + V_\delta \cdot \log_{10} f_\text{GHz} + V_\beta \cdot 
        \log_{10}\alpha_\text{rad} \right)\text{,}
\end{multline}
where ${X}$ is a Normal distributed random variables having zero-mean and unit variance. The seven parameters are: the reference value $V_\mu$ at 1~GHz, 1~m distance, and 1~rad (57.2\degree ) elevation; the distance dependence $V_\epsilon$ of the reference value scaling with $\log_{10}d$; the frequency dependence $V_\gamma$ of the reference value scaling with $\log_{10}f_\text{GHz}$; the elevation dependence $V_\alpha$ of the reference value scaling with $\log_{10}\alpha_\text{rad}$; the reference \ac{STD} $V_\sigma$ at 1~GHz, 1~m distance, and 1 rad elevation; the frequency dependence $V_\delta$ of the reference \ac{STD} scaling with $\log_{10}f_\text{GHz}$; and the elevation dependence $V_\beta$ of the reference \ac{STD} scaling with $\log_{10}\alpha_\text{rad}$. These parameters can be used in \acp{GSCM} to generate a randomized propagation environment for creating channel coefficients for simulation studies. 

The analysis results are shown in Table~\ref{tab:all_sat_parameters}. The last step is a resimulation step where the scatterers are generated by the QuaDRiGa channel model as described in \cite{3gpp_tr_38901_v1610}. Random delays and angles are generated based on the parameters from Table~\ref{tab:all_sat_parameters} and channel coefficients are created using the same antenna model. The evaluation procedure is repeated for these channel coefficients to confirm that the resulting parameters are similar to the ones in  Table~\ref{tab:all_sat_parameters}.

\begin{table*}
    \setlength{\tabcolsep}{1.5pt}
    \scriptsize
    \centering
    \caption{Large-Scale Parameters for Satellite Channel Models}\label{tab:all_sat_parameters}
\begin{tabular}{ llc | dd | dd | dd | dd | dd | dd  }

Parameter & Unit & 
    & \multicolumn{2}{c|}{Dense Urban}
    & \multicolumn{2}{c|}{Urban}
    & \multicolumn{2}{c|}{Suburban}
    & \multicolumn{2}{c|}{Rural}
    & \multicolumn{2}{c|}{\lt Dense Urb. Resim.} 
    & \multicolumn{2}{c}{\lt Rural Resim.} \\

&&& \multicolumn{1}{C{10mm}}{LOS} & \multicolumn{1}{C{10mm}|}{NLOS}
    & \multicolumn{1}{C{10mm}}{LOS} & \multicolumn{1}{C{10mm}|}{NLOS}
    & \multicolumn{1}{C{10mm}}{LOS} & \multicolumn{1}{C{10mm}|}{NLOS}
    & \multicolumn{1}{C{10mm}}{LOS} & \multicolumn{1}{C{10mm}|}{NLOS} 
    & \multicolumn{1}{C{10mm}}{\lt LOS} & \multicolumn{1}{C{10mm}|}{\lt NLOS}
    & \multicolumn{1}{C{10mm}}{\lt LOS} & \multicolumn{1}{C{10mm}}{\lt NLOS}\\

\hline\hline
No. clusters       & \textcolor[rgb]{0.75,0.75,0.75}{N/A} & $L$    &  11   &  10   &  11   &  10   &  9    &  8    & 7     &  6    &\lt 31   &\lt  30   &\lt  19   &\lt 18 \\
\cmidrule{1-15}
\textbf{PL}        & dB                      & $\text{PL}_\mu$     & 32.45 & 54.9 & 32.45 & 54.9  & 32.45 & 47.5  & 32.45 & 47.5  &\lt 31.1  &\lt 52.65 &\lt 31.05 &\lt 47.95  \\
PL dist. dep.      & dB / \lu{m}             & $\text{PL}_\epsilon$& 20.0  & 20.0 & 20.0  & 20.0  & 20.0  & 20.0  & 20.0  & 20.0  &\lt 20.05 &\lt 20.2  &\lt 20.0  &\lt 19.7 \\
PL freq. dep.      & dB / \lu{GHz}           & $\text{PL}_\gamma$  & 20.0  & 27.9 & 20.0  & 27.9  & 20.0  & 22.8  & 20.0  & 22.8  &\lt 20.95 &\lt 29.6  &\lt 20.95 &\lt 24.2 \\
PL elevation dep.  & dB / \lu{rad}           & $\text{PL}_\alpha$  &  \ze  &-11.0 &  \ze  &-11.0  &  \ze  & -8.4  &  \ze  & -8.4  &\lt -1.2  &\lt -11.05&\lt -1.3  &\lt -9.6 \\
Shadow Fading      & dB                      & $\text{SF}_\sigma$  &  0.15 & 10.0 &  0.1  &  6.0  &  1.45 & 10.4  &  1.4  & 10.1  &\lt 0.15  &\lt 10.2  &\lt  1.4  &\lt 9.6 \\
SF freq. dep.      & dB / \lu{GHz}           & $\text{SF}_\delta$  &  \ze  &  2.5 &  \ze  & \zel  &  \ze  & 0.75  &  \ze  &  1.1  &\zeg      &\lt  2.4  &\zeg      &\lt 1.45 \\
SF elev. dep.      & dB / \lu{rad}           & $\text{SF}_\beta$   & -0.6  & -2.5 &  \ze  & \zel  &  0.85 & 1.25  &  1.0  &  1.3  &\lt -0.85 &\lt -2.55 &\lt  0.7  &\lt 0.75 \\
SF decorr. dist.   & m                       & $\text{SF}_\lambda$ & 50.0  & 50.0 & 50.0  & 50.0  & 50.0  & 50.0  & 50.0  & 120.0 &\nag      &\nagl     &\nag      &\nag \\
\cmidrule{1-15}
\textbf{KF}        & db                      & $\text{KF}_\mu$     & 22.45& \nal  & 22.45 & \nal  & 13.95 & \nal  &  15.0 & \nal  &\lt 21.75 &\nagl     &\lt 14.35 &\nag \\
KF freq. dep.      & db / \lu{GHz}           & $\text{KF}_\gamma$  &  7.9 & \nal  &  7.9  & \nal  &  2.8  & \nal  &   2.8 & \nal  &\lt  8.15 &\nagl     &\lt 3.25  &\nag \\
KF elevation dep.  & db / \lu{rad}           & $\text{KF}_\alpha$  &-11.0 & \nal  & -11.0 & \nal  & -8.4  & \nal  &  -8.3 & \nal  &\lt -11.85&\nagl     &\lt -8.75 &\nag \\
KF STD             & db                      & $\text{KF}_\sigma$  & 10.6 & \nal  &  5.65 & \nal  & 11.05 & \nal  &  9.85 & \nal  &\lt 10.65 &\nagl     &\lt  9.55 &\nag \\
KF STD freq. dep.  & db / \lu{GHz}           & $\text{KF}_\delta$  &  2.2 & \nal  &  \ze  & \nal  &  \ze  & \nal  &  1.15 & \nal  &\lt  2.2  &\nagl     &\lt  1.4  &\nag \\
KF STD elev. dep.  & db / \lu{rad}           & $\text{KF}_\beta$   & -2.65& \nal  &  \ze  & \nal  &  1.3  & \nal  &  1.4  & \nal  &\lt -3.05 &\nagl     &\lt  1.05 &\nag \\
KF decorr. dist.   & m                       & $\text{KF}_\lambda$ & 50.0 & \nal  & 50.0  & \nal  & 50.0  & \nal  & 50.0  & \nal  &\nag      &\nagl     &\nag      &\nag \\
\cmidrule{1-15}
\textbf{DS}        & \lu{s}                  & $\text{DS}_\mu$     & -7.95 &-6.95 &  -7.8 & -6.85 & -7.45 & -6.7  & -6.85 & -6.1  &\lt -7.9  &\lt -6.95 &\lt -6.85 &\lt -6.1 \\
DS freq. dep.      & \lu{s} / \lu{GHz}       & $\text{DS}_\gamma$  & -0.4  & \zel &  -0.4 & \zel  &  \ze  & \zel  &  \ze  & \zel  &\lt -0.4  &\zegl     &\zeg      &\zeg  \\
DS elevation dep.  & \lu{s} / \lu{rad}       & $\text{DS}_\alpha$  &  0.4  & \zel &   0.5 & \zel  &  0.35 & \zel  & 0.35  & \zel  &\lt  0.45 &\zegl     &\lt 0.35  &\zeg  \\
DS STD             & \lu{s}                  & $\text{DS}_\sigma$  &  0.7  & 0.15 &   0.3 &  0.15 &  0.5  &  0.15 &  0.5  & 0.2   &\lt  0.7  &\lt 0.15  &\lt  0.5  &\lt 0.2 \\
Delay Factor &\textcolor[rgb]{0.75,0.75,0.75}{N/A} & $r_\text{DS}$ &  2.5  & 2.3  &  2.0  & 2.3   &  2.3  &  2.3  &  3.8  & 1.7   &\nag      &\nagl     &\nag      &\nag \\
Clst. DS           & ns                      & $\text{cDS}_\mu$    &  4.95 & 4.95 & 4.95  & 4.95  & 4.95  & 4.95  & 4.95  & 4.95  &\nag      &\nagl     &\nag      &\nag \\
Clst. DS freq. dep.& ns / \lu{GHz}           & $\text{cDS}_\gamma$ & -2.2  & -2.2 & -2.2  & -2.2  & -2.2  & -2.2  & -2.2  & -2.2  &\nag      &\nagl     &\nag      &\nag \\
DS decorr. dist.   & m                       & $\text{DS}_\lambda$ & 50.0  & 40.0 & 50.0  & 40.0  &  50.0 & 40.0  & 50.0  & 36.0  &\nag      &\nagl     &\nag      &\nag \\
\cmidrule{1-15}
\textbf{ASA}       & \lu{\degree}            & $\text{ASA}_\mu$    &  0.9  & 1.9  &  0.9  &  1.9  &  1.15 &  1.9  &  1.1  & 1.85  &\lt 0.8   &\lt 1.85  &\lt  1.1  &\lt 1.8  \\
ASA freq. dep.     & \lu{\degree} / \lu{GHz} & $\text{ASA}_\gamma$ & -0.4  & \zel & -0.4  & \zel  & \ze   & \zel  &  \ze  & \zel  &\lt -0.4  &\zegl     &\lt -0.1  &\zeg  \\
ASA elevation dep. & \lu{\degree} / \lu{rad} & $\text{ASA}_\alpha$ &  0.55 & \zel &  0.5  & \zel  &  0.4  & \zel  &  0.4  & \zel  &\lt 0.65  &\zegl     &\lt  0.45 &\zeg  \\
ASA STD            & \lu{\degree}            & $\text{ASA}_\sigma$ &  0.7  & 0.1  &  0.3  &  0.1  &  0.5  &  0.1  &  0.5  & 0.1   &\lt 0.7   &\lt 0.05  &\lt  0.5  &\lt 0.1 \\
Cluster ASA        & \degree                 & $\text{cASA}$       &  3.0  & 3.0  &  3.0  &  3.0  &  3.0  &  3.0  &  3.0  & 3.0   &\nag      &\nagl     &\nag      &\nag \\
ASA decorr. dist.  & m                       & $\text{ASA}_\lambda$& 50.0  & 50.0 &  50.0 & 50.0  & 50.0  & 50.0  & 50.0  & 40.0  &\nag      &\nagl     &\nag      &\nag \\
\cmidrule{1-15}
\textbf{ESA}       & \lu{\degree}            & $\text{ESA}_\mu$    & 0.3   & 1.25 &  0.45 & 1.0   &  0.85 &  0.4  &  0.8  & -0.4  &\lt 0.55  &\lt 1.25  &\lt  0.8  &\lt 0.1 \\
ESA freq. dep.     & \lu{\degree} / \lu{GHz} & $\text{ESA}_\gamma$ & -0.4  & \zel & -0.4  & \zel  &  \ze  & \zel  &  \ze  & \zel  &\lt -0.4  &\zegl     &\zeg      &\zeg  \\
ESA elevation dep. & \lu{\degree} / \lu{rad} & $\text{ESA}_\alpha$ & 0.7   & \zel &  1.15 & \zel  &  1.4  & \zel  &  1.4  & \zel  &\lt  1.0  &\zegl     &\lt  1.25 &\zeg  \\
ESA STD            & \lu{\degree}            & $\text{ESA}_\sigma$ & 0.7   & 0.15 &  0.3  & 0.25  &  0.5  &  0.4  &  0.5  & 0.5   &\lt  0.7  &\lt 0.15  &\lt  0.5  &\lt 0.2 \\
Cluster ESA        & \degree                 & $\text{cESA}$       & 1.0   & 1.0  & 1.0   & 1.0   &  1.0  &  1.0  &  1.0  & 1.0   &\nag      &\nagl     &\nag      &\nag \\
ESA decorr. dist.  & m                       & $\text{ESA}_\lambda$& 50.0  & 50.0 &  50.0 & 50.0  & 50.0  & 50.0  & 50.0  & 50.0  &\nag      &\nagl     &\nag      &\nag \\
\cmidrule{1-15}
\textbf{ASD}       & \lu{\degree}            & $\text{ASD}_\mu$    & 1.85  & 2.9  &  1.95 &  3.05 &  2.25 & 3.20  &   2.9 &  3.75 &\lt 1.75  &\lt 2.95  &\lt  3.0  &\lt 3.85 \\
ASD dist. dep.     & \lu{\degree} / \lu{m}  & $\text{ASD}_\epsilon$& -1.0  & -1.0 &  -1.0 & -1.0  &  -1.0 & -1.0  &  -1.0 & -0.95 &\lt -0.95 &\lt -1.0  &\lt -1.0  &\lt -0.95  \\
ASD freq. dep.     & \lu{\degree} / \lu{GHz} & $\text{ASD}_\gamma$ & -0.4  & \zel &  -0.4 & \zel  &  \ze  & \zel  &  \ze  & \zel  &\lt -0.45 &\zegl     &\zeg      &\zeg  \\
ASD elevation dep. & \lu{\degree} / \lu{rad} & $\text{ASD}_\alpha$ & 0.3   &-0.25 &  0.85 & -0.25 &  \ze  & -0.25 &  \ze  & -0.25 &\lt  0.35 &\lt -0.25 &\zeg      &\lt -0.25 \\
ASD STD            & \lu{\degree}            & $\text{ASD}_\sigma$ & 0.7   & 0.25 &  0.35 &  0.2  &  0.55 &  0.2  &  0.55 & 0.25  &\lt  0.7  &\lt  0.25 &\lt  0.5  &\lt 0.3  \\
ASD decorr. dist.  & m                       & $\text{ASD}_\lambda$& 50.0  & 50.0 & 50.0  & 50.0  & 50.0  &  50.0 & 50.0  & 30.0  &\nag      &\nagl     &\nag      &\nag \\
\cmidrule{1-15}
\textbf{ESD}       & \lu{\degree}            & $\text{ESD}_\mu$    & 1.75  & 2.85 &  1.8  &  2.95 &  2.15 & 3.1   &  2.85 & 3.7   &\lt  1.7  &\lt 2.90  &\lt  2.9  &\lt 3.75 \\
ESD dist. dep.     & \lu{\degree} / \lu{m}  & $\text{ESD}_\epsilon$& -1.0  & -1.0 &  -1.0 & -1.0  & -0.95 & -1.0  & -1.0  & -1.0  &\lt  -1.0 &\lt -1.0  &\lt -1.0  &\lt -1.0 \\
ESD freq. dep.     & \lu{\degree} / \lu{GHz} & $\text{ESD}_\gamma$ & -0.4  & \zel &  -0.4 &  \zel &  \ze  & \zel  &  \ze  & \zel  &\lt -0.4  &\zegl     &\zeg      &\zeg  \\
ESD elevation dep. & \lu{\degree} / \lu{rad} & $\text{ESD}_\alpha$ & 0.5   & \zel &  0.85 &  0.5  &  1.05 & 0.65  &  1.05 & 0.65  &\lt  0.5  &\zegl     &\lt  1.1  &\lt 0.65 \\
ESD STD            & \lu{\degree}            & $\text{ESD}_\sigma$ & 0.7   & 0.15 &  0.35 &  0.2  &  0.55 & 0.2   &  0.55 & 0.2   &\lt  0.7  &\lt  0.15 &\lt  0.55 &\lt 0.25 \\
ESD decorr. dist.  & m                       & $\text{ESD}_\lambda$& 50.0  & 50.0 & 50.0  & 50.0  & 50.0  &  50.0 & 50.0  & 50.0  &\nag      &\nagl     &\nag      &\nag \\
\cmidrule{1-15}
\textbf{XPR}       & db                      & $\text{XPR}_\mu$    & 15.15 & 15.15&  7.0  &  7.0  & 12.65 & 12.65 &  7.0  &  7.0  &\lt 14.75 &\lt 14.4  &\lt  7.0  &\lt 7.05 \\
XPR elevation dep. & db / \lu{rad}           & $\text{XPR}_\alpha$ & -13.45&-13.45&  \ze  &  \zel &-11.2  &-11.2  &  \ze  &  \zel &\lt -13.5 &\lt -14.2 &\zeg      &\zeg  \\
XPR STD            & db                      & $\text{XPR}_\sigma$ & 13.65 & 13.65&  3.0  &  3.0  & 10.95 & 10.95 &  3.0  &  3.0  &\lt 13.55 &\lt  12.7 &\lt 3.0   &\lt 2.95 \\
XPR STD el. dep.   & db / \lu{rad}           & $\text{XPR}_\beta$  &  8.8  & 8.85 &  \ze  &  \zel &  2.7  &  2.7  &  \ze  &  \zel &\lt  9.45 &\lt  9.25 &\zeg      &\zeg  \\
XPR decorr. dist.  & m                       & $\text{XPR}_\lambda$& 50.0  & 50.0 & 50.0  & 50.0  & 50.0  & 50.0  & 50.0  & 40.0  &\nag      &\nagl     &\nag      &\nag \\
\end{tabular}
\end{table*}

\section{Results and Discussion}\label{sec:lsf-model-results}

The fitted parameters for the eight \acp{LSP} are shown in Table~\ref{tab:all_sat_parameters}, together with the resimulation results for the dense urban and rural scenarios. The inter-parameter correlation values of the random variables $X$ in \eqref{eq:lsf_linear_model} are given in Table~\ref{tab:Sat-Cross-Corr}. The upper right part (shown in white) contains the values for the \ac{LOS} channels, the lower left part shows the values for the \ac{NLOS} channels. Our simplified propagation model doses not consider polarization effects which need detailed information about the materials and incidence angles. Hence, we reuse the \ac{XPR} values from \cite{3gpp_tr_38811_v1520} and fit the results to \eqref{eq:lsf_linear_model}. In the following, the results are discussed:

\paragraph*{\Acf{PL}} In \cite{3gpp_tr_38811_v1520}, the \ac{NLOS}-\ac{PL} is modeled by the \ac{FSPL} and additional clutter loss, which models the attenuation caused by buildings and objects on the ground. We used this model for the \ac{NLOS} parameters in Table~\ref{tab:all_sat_parameters}. Depending on the scenario, the elevation angle and the frequency, the \ac{NLOS}-\ac{PL} is about 15 to 45~dB higher compared to the \ac{FSPL}. This leads to very low received \ac{NLOS} power, e.g.\ in urban or dense urban scenarios where the clutter loss is highest. When adding a \ac{LOS} component (assuming that the \ac{NLOS} power does not change), it dominates the overall \ac{PL} formula since it has about 30 to 3000 times more power compared to the scattered signals from buildings and vegetation. This is also the case in \cite{3gpp_tr_38811_v1520} where the \ac{LOS}-\ac{PL} is the same as the \ac{FSPL}. In the resimulation, the parameterized model is able to create similar values. However, there is a slight increase in the frequency-dependence $\text{PL}_\gamma$  which is compensated by a slightly lower base value $\text{PL}_\mu$. 

\paragraph*{\Acf{SF}} \Ac{SF} occurs when an obstacle gets positioned between the satellite and the \ac{MT}. This leads to a reduction in signal strength because the wave is shadowed or blocked by the obstacle. We used the \ac{NLOS}-\ac{SF} model parameters from \cite{3gpp_tr_38811_v1520} to model the power fluctuations of the scattered paths. However, the power of the \ac{LOS} component is deterministic. Hence, the \ac{LOS}-\ac{SF} is much smaller compared to the \ac{NLOS}-\ac{SF} because only the scattered paths can vary in strength. This is the case in our results in Table~\ref{tab:all_sat_parameters}. However, 
\cite{3gpp_tr_38811_v1520} reports significantly larger \ac{LOS}-\ac{SF} of 4 dB for the urban and dense urban scenarios. Partial shadowing of the first Fresnel zone might cause fluctuations of the \ac{LOS} path strength which increases the \ac{SF}.

\paragraph*{\Acf{KF}} The \ac{KF} is the ratio between the power of the direct path and the power of scattered paths. It is only defined for \ac{LOS} scenarios. Due to our characterization method, the \ac{KF} in Table~\ref{tab:all_sat_parameters} reflects the difference between the \ac{NLOS}-\ac{PL} and the \ac{FSPL}. Values can range up to 40~dB in the urban and dense urban scenarios at low elevation in the KA-band. The values reported in \cite{3gpp_tr_38811_v1520} are consistent with our findings only for the urban and the rural scenario. The dense urban and the suburban show show much lower KF results, especially at low elevation angles. It is unclear where this inconsistency comes from.

\paragraph*{\Acf{DS}} The \ac{DS} is an important measure for the delay time extent of a multipath radio channel. It is defined as the square root of the second central moment of the power-delay profile. The \ac{NLOS}-\ac{DS} is consistent with the distance to the scatterers in Table~\ref{tab:propagation_environment}. The dense urban scenario has an average value of 112~ns. This value increases with decreasing building density to 794~ns in the rural scenario. There is no frequency or elevation dependence in our model, since local scattering does not depend on the satellite position. In contrary, values reported in \cite{3gpp_tr_38811_v1520} show a decreasing \ac{DS} for increasing $\alpha$ in the three urban scenarios and increasing \ac{DS} for the rural scenario. The reported values are much smaller at 41~ns in the dense urban scenario and 11~ns for the rural scenario ($\alpha$ = 50\degree , $f$ = 20 GHz). This would mean that in the rural scenario, the average distance between the \ac{MT} and the scattering objects is only about 3~m compared to the 240~m in our model. When a \ac{LOS} component is added to the existing \ac{NLOS} model, the \ac{DS} decreases since most power is now allocated to the direct path. The \ac{LOS}-\ac{DS} parameters in Table~\ref{tab:propagation_environment} also inherit the frequency and elevation dependence from the \ac{KF} and there is a strong negative correlation between the \ac{LOS}-\ac{DS} and the \ac{KF} in Table~\ref{tab:Sat-Cross-Corr}. These strong correlations are not reported by \cite{3gpp_tr_38811_v1520} and the effect of the \ac{KF} on the \ac{DS} is not as strong. The resimulation results show an almost perfect match with the initial parameters. However, the per-cluster \ac{DS} introduced by \cite{3gpp_tr_38811_v1520} effectively splits each cluster into three sub-clusters which tipples the number of clusters in the resimulation output.

\begin{table}[t]
   \setlength{\tabcolsep}{1.8pt}
    \scriptsize
    \centering
    \caption{Inter-Parameter Correlation Values}\label{tab:Sat-Cross-Corr}
    \vspace{-2mm}
\begin{tabular}{ lll | ddddddd }
 \multicolumn{3}{l|}{ Inter-Parameter } & \multicolumn{7}{c}{L O S} \\
 \multicolumn{3}{l|}{ Correlations  }
                     & \tbt{DS} & \tbt{KF} & \tbt{SF} & \tbt{ASD} & \tbt{ASA} & \tbt{ESD} & \tbt{ESA} \\
 \hline \hline
 & DS  & Dense Urban & \tp 1    & -0.8     &  0.2     &  0.8      &   0.8     &  0.8      & 0.8 \\
 &     & Urban       & \tp 1    & -0.8     &  0.1     &  0.8      &   0.8     &  0.8      & 0.8 \\
 &	   & Suburban    & \tp 1    & -0.8     &  0.4     &  0.8      &   0.8     &  0.8      & 0.8 \\
 &     & Rural       & \tp 1    & -0.8     &  0.4     &  0.8      &   0.8     &  0.8      & 0.8 \\
\cmidrule(l){2-10}
 & KF  & Dense Urban & \lo  0   & \tp 1    & -0.3     & -0.8      &  -0.8     & -0.8      & -0.8 \\
 &     & Urban       & \lo -0.1 & \tp 1    & -0.1     & -0.8      &  -0.8     & -0.8      & -0.8 \\
 &	   & Suburban    & \lo  0   & \tp 1    & -0.6     & -0.8      &  -0.8     & -0.8      & -0.8 \\
 &     & Rural       & \lo  0   & \tp 1    & -0.5     & -0.8      &  -0.8     & -0.8      & -0.8 \\
\cmidrule(l){2-10}
 & SF  & Dense Urban & \lo  0   & \lona    & \tp 1    &  0.2      &   0.2     &  0.2      & 0.2 \\
N&     & Urban       & \lo  0   & \lona    & \tp 1    &  0.1      &   0.1     &  0.1      & 0.1 \\
 &	   & Suburban    & \lo  0   & \lona    & \tp 1    &  0.5      &   0.5     &  0.5      & 0.4 \\
L&     & Rural       & \lo -0.1 & \lona    & \tp 1    &  0.5      &   0.5     &  0.4      & 0.4 \\
\cmidrule(l){2-10}
O& ASD & Dense Urban & \lo  0.2 & \lona    & \lo  0   & \tp 1     &   0.8     &  0.8      & 0.8 \\
 &     & Urban       & \lo  0.2 & \lona    & \lo  0   & \tp 1     &   0.8     &  0.8      & 0.7 \\
S&	   & Suburban    & \lo  0.2 & \lona    & \lo  0   & \tp 1     &   0.8     &  0.8      & 0.8 \\
 &     & Rural       & \lo  0.2 & \lona    & \lo  0   & \tp 1     &   0.8     &  0.8      & 0.8 \\
\cmidrule(l){2-10}
 & ASA & Dense Urb.  & \lo  0.1 & \lona    & \lo  0   & \lo  0.1 & \tp 1      &  0.8      & 0.8 \\
 &     & Urban       & \lo  0.1 & \lona    & \lo  0   & \lo  0.1 & \tp 1      &  0.8      & 0.8 \\
 &	   & Suburban    & \lo  0.1 & \lona    & \lo  0   & \lo  0.1 & \tp 1      &  0.8      & 0.8 \\
 &     & Rural       & \lo  0.2 & \lona    & \lo  0   & \lo  0.2 & \tp 1      &  0.8      & 0.8 \\
\cmidrule(l){2-10}
 & ESD & Dense Urb.  & \lo  0.1 & \lona    & \lo  0   & \lo  0.3 & \lo  0.1   & \tp 1     & 0.8 \\
 &     & Urban       & \lo  0.4 & \lona    & \lo  0   & \lo  0.4 & \lo  0.1   & \tp 1     & 0.8 \\
 &	   & Suburban    & \lo  0.5 & \lona    & \lo  0   & \lo  0.3 & \lo  0.2   & \tp 1     & 0.8 \\
 &     & Rural       & \lo  0.5 & \lona    & \lo  0   & \lo  0.3 & \lo  0.2   & \tp 1     & 0.8 \\
\cmidrule(l){2-10}
 & ESA & Dense Urb.  & \lo -0.1 & \lona    & \lo  0   & \lo -0.1 & \lo  0     & \lo  0    & \tp 1 \\
 &     & Urban       & \lo -0.1 & \lona    & \lo  0   & \lo -0.2 & \lo  0     & \lo -0.1  & \tp 1 \\
 &	   & Suburban    & \lo -0.1 & \lona    & \lo  0   & \lo -0.2 & \lo  0     & \lo -0.2  & \tp 1 \\
 &     & Rural       & \lo  0   & \lona    & \lo  0   & \lo -0.2 & \lo  0     & \lo -0.2  & \tp 1 \\
\end{tabular}
\end{table}

\paragraph*{\Acf{ASA}}
In our model, the \ac{ASA} was drawn from a Uniform distribution in the range $[-\pi,\pi[$. Given that the scattered paths have similar power values, the average \ac{NLOS}-\ac{ASA} for 10 clusters is 84\degree . This value is achieved consistently for all scenarios in our model. However, values in \cite{3gpp_tr_38811_v1520} indicate much smaller \ac{NLOS}-\acp{ASA} of around 30\degree\ for the dense urban scenario and 3-6\degree\ for the other scenarios. At the same time, \cite{3gpp_tr_38811_v1520} proposes a large per-cluster \ac{ASA} of up to 30\degree\ and a small number of 2-4 clusters. Defining an \ac{AS} for only 2 clusters while at the same time having a large \ac{AS} within the clusters seems unreasonable. There is also no description of the clustering method nor an explanation and discussion of the results, making it hard to interpret the findings. As for the \ac{DS}, the \ac{ASA} decreases when a \ac{LOS} component is added. A frequency and elevation dependence of the \ac{LOS}-\ac{ASA} can also be observed due to the negative correlation with the \ac{KF}. 

\paragraph*{\Acf{ESA}}
Since the average building height decreases in our model when moving from a dense urban to a rural scenario, the \ac{NLOS}-\ac{ESA} decreases as well from around 18\degree\ to well below 1\degree . When we add a \ac{LOS} component, the \ac{ESA} values decrease for the Urban and increase for the Rural scenarios. The scattered paths always arrive from the horizontal plane. Hence, there is an increasing difference in the elevations components between the \ac{NLOS} paths and the direct path. This increases the \ac{ESA} when the satellite is high up in the sky. All values reported in \cite{3gpp_tr_38811_v1520} have a strong elevation-dependency. This intuitively makes sense for the \ac{LOS} channels where the direct path has a strong influence on the \ac{ESA}. However, there is no explanation for the \ac{NLOS} channels. For example, in the Rural-\ac{NLOS} scenario, the \ac{ESA} changes from 0.1\degree\ when the satellite is at the horizon to 22\degree\ when it is at the zenith. Calculating scatterer positions from those angles places them several kilometer up in the air which make no sense in a Rural setting with mostly low buildings. Also, \cite{3gpp_tr_38811_v1520} reports a strong frequency-dependence in the suburban case where at $\alpha$ = 50\degree\ the S-band \ac{ESA} is 0.01\degree\ and the KA-band value is 24\degree . This is not reported for the other scenarios.

\paragraph*{\Acf{ASD} and \acf{ESD}}
The departure-\acp{AS} plays an important role in multi-beam satellite systems. The model parameters must reflect the multipath environment on the ground. Too large values might cause interference in neighboring beams. However, the beam centers are often separated by several hundred kilometers on the ground, outside the range of scattered paths. Since the satellite can be in different orbit heights, \ac{ASD} and \ac{ESD} must depend on the distance between the satellite and the terminal. This is the case for the values in Table~\ref{tab:all_sat_parameters}, where the departure-\acp{AS} decrease with increasing distance. It is not considered by \cite{3gpp_tr_38811_v1520}. In addition to the distance-dependence, there is a dependence on the satellite elevation angle. The \ac{ESD} is smallest when the satellite is just above the horizon and increases when it moves to the zenith. This is also not reported by \cite{3gpp_tr_38811_v1520}.

\section{Conclusions}

In this paper, we proposed a non-geostationary-satellite motion model which has been integrated into the open-source \ac{QuaDRiGa} channel model. We also proposed a simplified method to obtain the model parameters and we were able to produce a consistent set of parameters for four typical environments. Our parameter-set also produces almost identical calibration results as the parameter-set provided by 3GPP \cite{3gpp_tr_38821_v1600}\footnote{The calibration was done using QuaDRiGa v2.4. The source code and the results are available at \cite{quadriga_www}}. This is mainly due to the fact, the calibration was done for \ac{LOS} channels only which are dominated by the direct path. However, our method does not include many important radio propagation effects such as diffraction, polarization or interactions with different kinds of materials. It is therefore important to use measurements or ray-tracing methods to refine these parameters in the future. 

\bibliographystyle{IEEEtran}
\bibliography{ms} 

\end{document}